\documentclass{PoS}
\usepackage{rotating}
\usepackage{multirow}
\usepackage[authoryear,square]{natbib}
\bibpunct{(}{)}{;}{a}{}{,}

\title{The MWA GLEAM 4 Jy sample; a new large, bright radio source sample at 151 MHz}

\ShortTitle{The MWA GLEAM 4Jy sample}

\author{\speaker{C.A.Jackson}$^{1}$ T.M.O. Franzen${^1}$, N. Seymour${^1}$, S.V. White${^1}$, 
Tara Murphy$^{2,3}$, E. M. Sadler$^{2,3}$, J. R. Callingham$^{2,3,4}$, R. W. Hunstead${^2}$, J. Hughes$^{2}$, J. V. Wall${^8}$, M. E. Bell$^{3,4}$, K.S. Dwarakanath${^5}$, B-Q. For${^6}$, B.M. Gaensler$^{2,3,7}$,
P. J. Hancock$^{1,3}$, L. Hindson${^9}$, N. Hurley-Walker${^1}$, M. Johnston-Hollitt${^9}$,
A. D. Kapi\'{n}ska$^{3,6}$, E. Lenc$^{2,3}$, B. McKinley$^{10}$, J. Morgan${^1}$, 
A. R. Offringa$^{11}$, P. Procopio$^{10}$, L. Staveley-Smith$^{3,6}$
R. B. Wayth$^{1,3}$, C. Wu${^6}$, Q. Zheng${^9}$ \\

 $^{1}$ICRAR Curtin University, Australia; $^{2}$University of Sydney, Australia; $^{3}$ARC Centre of Excellence For All-Sky Astrophysics (CAASTRO) $^{4}$CSIRO Astronomy and Space Science (CASS), Australia; $^{5}$Raman Research Institute,  India; 
 $^{6}$ICRAR University of Western Australia, Australia; $^{7}$Dunlap Institute for Astronomy \& Astrophysics, University of Toronto, Canada;
 $^{8}$University of British Columbia, Canada; $^{9}$Victoria University of Wellington, New Zealand;
 $^{10}$The University of Melbourne, Australia; 
 $^{11}$Netherlands Institute for Radio Astronomy (ASTRON), The Netherlands \\
       Email: \email{carole.jackson@curtin.edu.au}}

\abstract{This paper outlines how the new GaLactic and Extragalactic All-sky MWA Survey (GLEAM, Wayth et al. 2015), observed by the Murchison Widefield Array covering the frequency range 72 $-$ 231 MHz, allows identification of a new large, complete, sample of more than 2000 bright extragalactic radio sources selected at 151 MHz.  With a flux density limit of 4~Jy this sample is significantly larger than the canonical fully-complete sample, 3CRR (Laing, Riley \& Longair 1983). In analysing this small bright subset of the GLEAM survey we are also providing a first-user check of the GLEAM catalogue ahead of its public release (Hurley-Walker et al. in prep). Whilst significant work remains to fully characterise our new bright source sample, in time it will provide important constraints to evolutionary behaviour, across a wide redshift and intrinsic radio power range, as well as being highly complementary to results from targeted, small area surveys.}  

\FullConference{EXTRA-RADSUR2015 (*)\\
		20--23 October 2015\\
		Bologna, Italy

                \bigskip
                \hrule
                \bigskip

                \textnormal{(*) This conference has been organized
                  with the support of the Ministry of Foreign Affairs
                  and International Cooperation, Directorate General
                  for the Country Promotion (Bilateral Grant Agreement
                  ZA14GR02 - Mapping the Universe on the Pathway to
                  SKA)}
}

\begin{document}

\section{Introduction}

Complete samples of radio sources are an essential tool to unravelling the cosmic evolution of radio galaxies and quasars. Whilst deep, small area surveys can be designed to detect sources of relatively low intrinsic radio power at moderate-to-high redshifts, they fail to detect statistically meaningful samples of the brightest and rarest sources. As a result we are driven to sample the largest accessible volume to trace the high-power, tail end of the radio luminosity function (RLF). However, any bright flux density-limited sample includes a wide diversity of sources beyond the most luminous and most distant; sources also lie at local and intermediate distances and thus it is necessary to obtain reliable distance measures to untangle the effects of luminosity and redshift. This requirement has long been recognised: the effort required to completely identify relatively modest scale radio samples has been enormous, although with the advent of all-sky multi-band imaging and (photometric) spectroscopy, this situation is improving. 

As this conference looks towards new scientific challenges, we note that deep extragalactic radio surveys are being used to define the scientific potential of future radio telescopes. The path to defining a new telescope is to ensure its capabilities are suitably matched to its sensitivity, e.g. considering the number density of sources, dynamic range, confusion effects, etc.  One approach is to extrapolate the best available sky model(s) to approximate the sensitivity of the proposed instrument. This is how the radio astronomical community is exploring the scientific potential of the Square Kilometre Array (SKA) telescope particularly in the continuum domain (Prandoni \& Seymour 2015), via extrapolation of existing source counts and other complementary data. The new series of deep, low frequency radio surveys such as the GaLactic and Extragalactic All-sky MWA Survey (GLEAM, Wayth et al. 2015) are vital in providing data in the same frequency range as SKA. 

\section{Challenges of evolution models and source counts}

Radio source counts have been measured from deep and increasingly wide-area surveys to very sensitive flux density limits of a few tens of $\mu$Jy, at frequencies between 1 - 3 GHz - a range now referred to as part of the `mid' frequency band within the SKA project.  At these observing frequencies, some fraction of sources appear `beamed', i.e. boosted in flux density via relativistic aberration from the physical alignment of core-jet structures close to our line-of-sight. Moreover, as radio sources lie at all redshifts, we observe intrinsic (rest-frame) emission from increasingly more compact and energetic regions of each radio source as both observing frequency and redshift increase. At mid frequencies the combined effect of these features is that we lose the ability to detect the intrinsic origin of the radio emission and instead only account for a decreasing fraction of the sources' radio activity. These two effects bias the source count and are the origin of the changing form of the differential source count with frequency noted by Wall (1994): the consequence is that it is not straightforward to extrapolate source counts to a significantly different observing frequency without knowledge of the intrinsic emission properties of the source population(s) plus the evolution of the RLF. 

Whilst we have fair insight into the evolution of the RLF for the extreme high-power sources from the 3CRR sample, it is not  determined for sources of moderate or low intrinsic radio power (e.g. $P_{151 \rm{MHz}} < 10^{25}$ W Hz$^{-1}$). There is little information on the space density of these radio source population(s) except locally (i.e. at $z < 0.1$): here some progress has been made using wide area radio surveys combined with sizable spectroscopic optical surveys (e.g. Mauch \& Sadler 2007). 

Within the context of defining SKA science, much work has been done to model the SKA sky to deep flux density limits (e.g. S$^{3}-$SEX, Wilman et al. 2008) taking account of the canonical `radio-loud' source populations and including source populations more normally described as `radio-quiet'. However due to the effects noted above, it is not trivial to translate these 1.4~GHz model skies to lower frequencies. This is particularly pertinent to the current SKA$\_$LOW specification (frequency range of 50 - 350~MHz) as has been illustrated by Franzen et al. (2016) and others. 

We have posited (Wall \& Jackson 1997) that at low frequencies ($\nu < 200$~MHz) the effects of beaming are negligible such that we observe unbiased source emission, for which we adopt the term the `parent' population(s). The evolutionary behaviour of each source population can then be transposed to higher frequencies by a simple model of jet Doppler factors and randomised sky orientation. This approach has been taken in whole or part by a number of authors, e.g. Orr \& Browne 1982; Wall \& Peacock 1985; Morisawa \& Takahara 1987; Wall \& Jackson 1997; Jackson \& Wall 1999, and has been reasonably successful in reproducing radio source counts across a wide frequency range (150~MHz - 5~GHz).  

\begin{figure}
\centering
\includegraphics[scale=0.30]{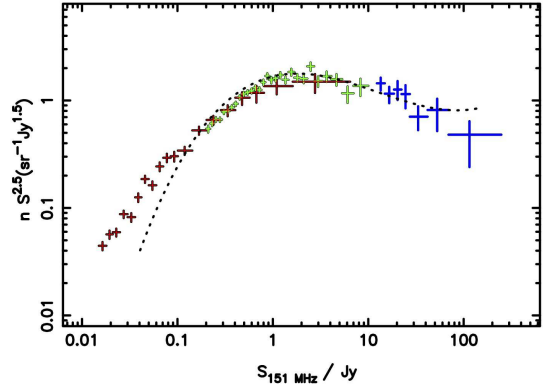}
\caption{Differential radio source counts at 151 MHz with the model fit of Jackson \& Wall (1999) shown dotted; the model diverges from the observed source counts below $S_{151 \rm{MHz}} \sim$ 0.1 Jy. Source count data is compiled from 3CRR (178~MHz: Laing, Riley \& Longair 1983) transposed to 151 MHz for $S>$ 12.33 Jy; 6C (151 MHz: Hales, Baldwin \& Warner 1988) over the range 0.2 Jy to 10 Jy and from the T-RaMiSu survey from 0.02 Jy to 6 Jy (153 MHz: Williams, Intema \& Rottgering 2013).}
\label{fit151f}
\end{figure}

Differential source counts from contemporary deep low frequency surveys now reveal that our model, based on extrapolation of the RLF derived from the 3CRR sample, underestimate the observed source count as shown in Figure \ref{fit151f}. We note that this mismatch had been noted when the first deep, small area low frequency source counts at 153 MHz were determined  (Intema et al. 2011).  Whilst such models were able to fit the 3CRR, 6C and 7C survey counts, it is now clear that the derived RLF is deficient in predicting the lower flux density count traced by wide field surveys such as the T-RaMiSu survey (Williams et al. 2013).  This shortfall could be due to a paucity of low redshift, lower power sources, or high redshift, high power sources or a combination of both. A path to resolving this degeneracy is to investigate the luminosity distribution of a large complete sample of sources of high flux density to trace the distribution of moderate-to-high power sources at much higher statistical significance than can be gleaned from 3CRR. 

\section{Low frequency surveys - old and new}
A number of major wide-area radio surveys have been conducted at frequencies between $\sim$100 MHz and 200 MHz, e.g. the Cambridge surveys 3C (159 MHz: Edge et al. 1959), 4C (178 MHz: Bennett 1962), 6C (151 MHz: Pilkington \& Scott 1965; Gower, Scott \& Wills 1967) and 7C (151 MHz: Hales et al. 1988, 2007). There are also more recent, ongoing surveys including the MRT survey (151.5 MHz: Pandey \& Shankar 2007), TGSS, a new deep survey at 153 MHz (Sirothia et al. 2010), the LOFAR surveys including MSSS (Heald et al. 2015) and the MWA GLEAM survey (Wayth et al. 2015). Whilst all of these surveys have revealed the complexity of the radio source sky, the number and sample sizes of complete, or near-completely-identified, samples has remained small as noted in Table~1: this is due to the enormous amount of effort required to locate and identify the hosts of the radio sources.  At slightly higher frequencies, samples such as the Molonglo Southern 4 Jy sample ("MS4" selected at 408 MHz, Burgess \& Hunstead 2006) provide comparable samples to 3CRR: whilst remaining limited to small numbers of sources, these samples also become increasingly susceptible to the observational bias effects described earlier.   

\begin{table}
 \centering
  \caption{Complete or near-completely identified low frequency ($\nu < 200$~MHz) radio source samples. }
   \label{smallsamps}
    \begin{tabular}{lccrrl}
\hline
\multicolumn{1}{c}{}& \multicolumn{1}{c}{Sample}
& \multicolumn{1}{c}{} 
& \multicolumn{1}{c}{Flux density}& \multicolumn{1}{c}{Optical/z}
& \multicolumn{1}{c}{} \\
\multicolumn{1}{c}{Survey}& \multicolumn{1}{c}{size} 
& \multicolumn{1}{c}{Frequency} 
& \multicolumn{1}{c}{limit} & \multicolumn{1}{c}{completeness}
& \multicolumn{1}{c}{Reference}\\
\hline
3CRR   & 173 & 178 MHz & 10.9 Jy & 100\% & Laing, Riley, Longair 1983 \\
7CI    &  37 & 151 MHz & 0.51 Jy & 90\% &  Grimes, Rawlings, Willott 2004 \\
7CII   &  54 & 151 MHz & 0.48 Jy     &  90\% & Grimes, Rawlings, Willott 2004 \\
7CIII   & 37 & 151 MHz & 0.5 Jy &  95\% & Lacy et al. 1999  \\
TOOTS-00 & 47  & 178 MHz & 0.1 Jy & $\sim$80\% & Vardoulaki et al. 2010 \\
\hline
\end{tabular}
\end{table}

This situation improves with the advent of multi-frequency all-sky radio surveys where low frequency, low resolution, radio data can be matched to higher frequency radio data to provide refined position estimates to confidently identify the host galaxy. These revised positions can be used across the EM spectrum, i.e. taking advantage of wide-field multi-passband optical imaging, spectroscopy, IR surveys, etc. However, we are also painfully aware that a sizable fraction of bright radio sources are found to have faint optical hosts; whilst this new era of wide spectrum coverage is upon us, it is likely that many will require singular follow-up in order to obtain a completely-identified sample.

\section{The MWA GLEAM 4 Jy sample}

The Murchison Widefield Array (MWA, Tingay et al. 2013, Ord et al. 2015) is sited at the Murchison Radio-astronomy Observatory (MRO) in Western Australia. The MWA has a range of science goals (Bowman et al. 2013), including the GLEAM survey (Wayth et al. 2015). This survey covers the sky in the declination range $+$25$^\circ$ to $-$80$^\circ$ with a near contiguous frequency coverage from 72 to 231 MHz.  

The GLEAM survey catalogue is constructed from a series of 120~s drift-scan observations, each set covering one of five, 30.72 MHz-wide instantaneous frequency bands between 72 and 231 MHz. From these observations, the GLEAM data are imaged and calibrated in 20, 7.68~MHz frequency sets, such that we can choose relatively narrow frequency slices to select any specific sample. A catalogue of approximately 300,000 extragalactic source components detected by the GLEAM survey during its first year of observations is nearing release (Hurley-Walker et al., in prep). This will provide full details of the flux density calibration procedures, completeness and reliability of the bright sample discussed further in this paper. 

A large fraction of the extragalactic GLEAM survey ($\sim$ 7.4 sr) is 100$\%$ complete at $S_{151 \rm{MHz}} \ge 4$ Jy excluding regions close to the Galactic Plane, $|b| < 10^\circ$ and Magellanic Clouds. From these data we select 2130 components with $S_{151 \rm{MHz}} \ge 4$ Jy catalogued in the GLEAM 147 - 154 MHz image.

\section{Reconciling the GLEAM catalogue to discrete radio sources}

As with almost all other extragalactic radio source catalogues, the catalogued GLEAM components do not necessarily have a one-to-one correspondence with physically discrete radio galaxies and quasars. How the GLEAM survey catalogue represents the physical sky in this respect is one of the first issues we need to resolve before attempting to interpret the GLEAM 4 Jy `sources' with any other radio data.  

\begin{figure}
\centering
\includegraphics[scale=0.25]{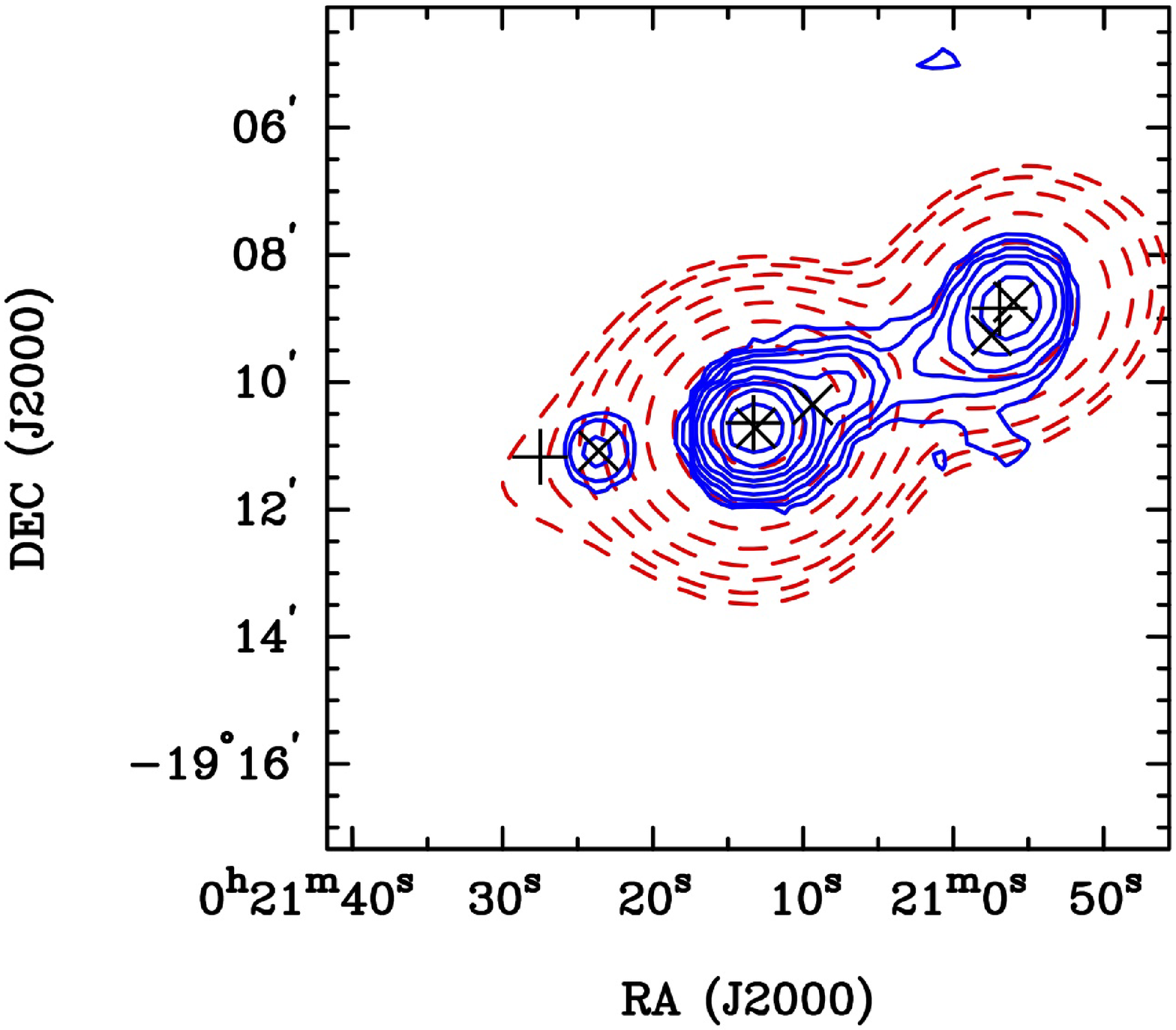}
\includegraphics[scale=0.25]{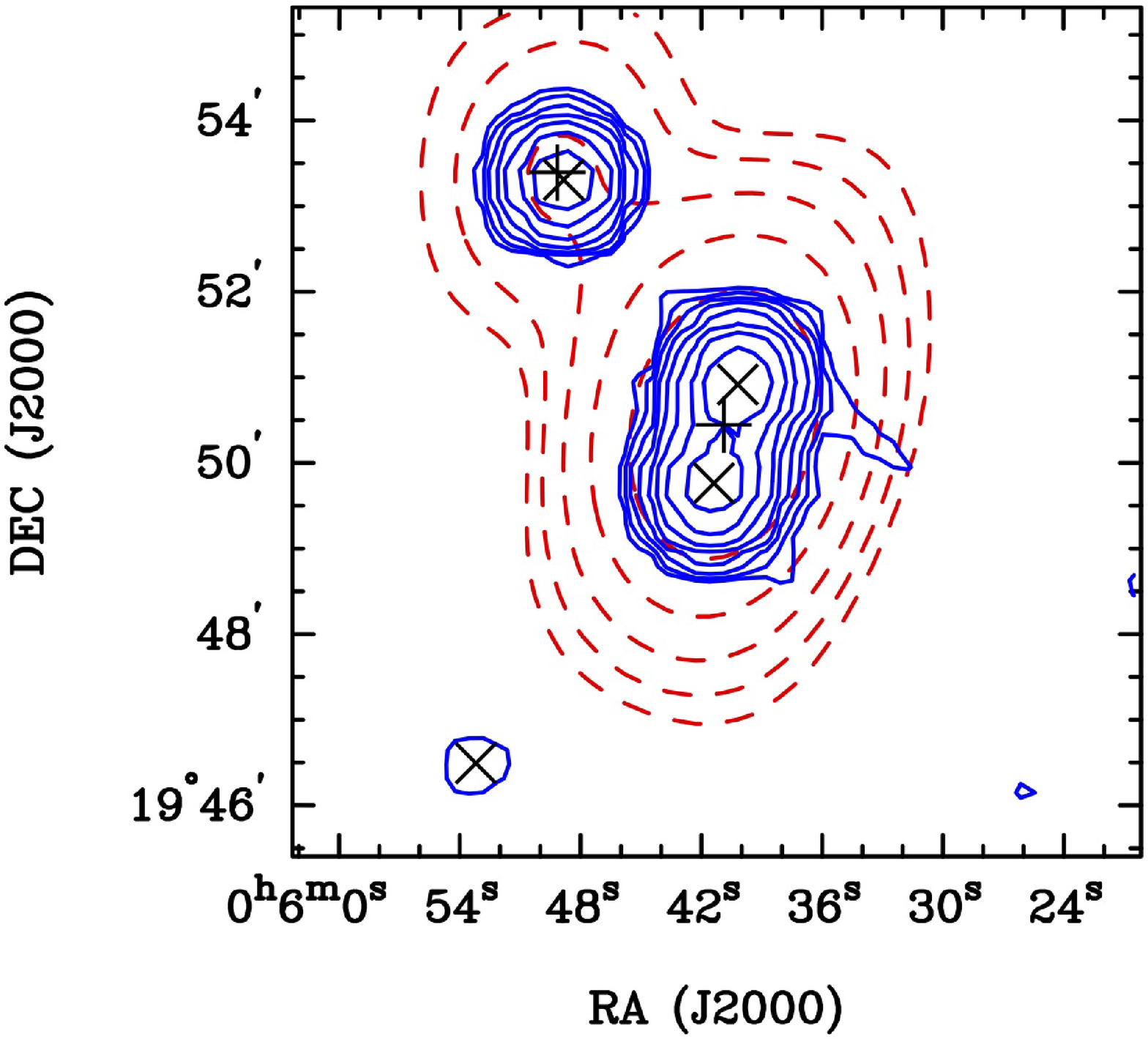}
\caption{An example of a resolved GLEAM double (left) and where multiple (in this case, two) GLEAM components are determined to be unrelated (right): sources GLEAM J002113-191038 and GLEAM J000540+195026 respectively. Dashed contours (GLEAM) and solid contours (NVSS) are shown with the lowest contour level at 3$\sigma$ (rms, survey) and increasing in factors of 2. The catalogued radio positions are shown for both surveys, GLEAM $+$ and NVSS $\times$.}
\label{gdoublea}
\end{figure}

\begin{figure}
\centering
\includegraphics[scale=0.25]{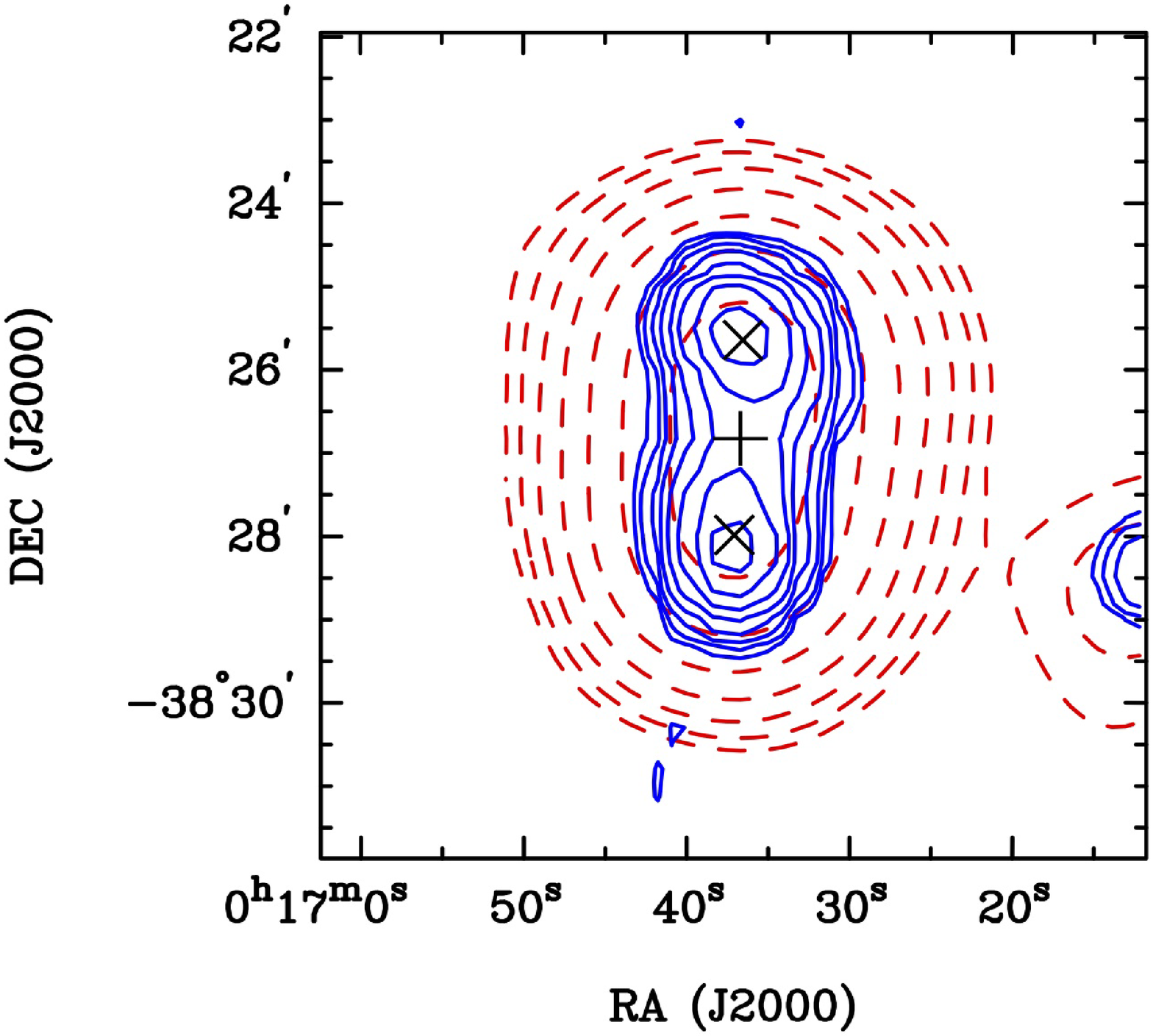}
\includegraphics[scale=0.25]{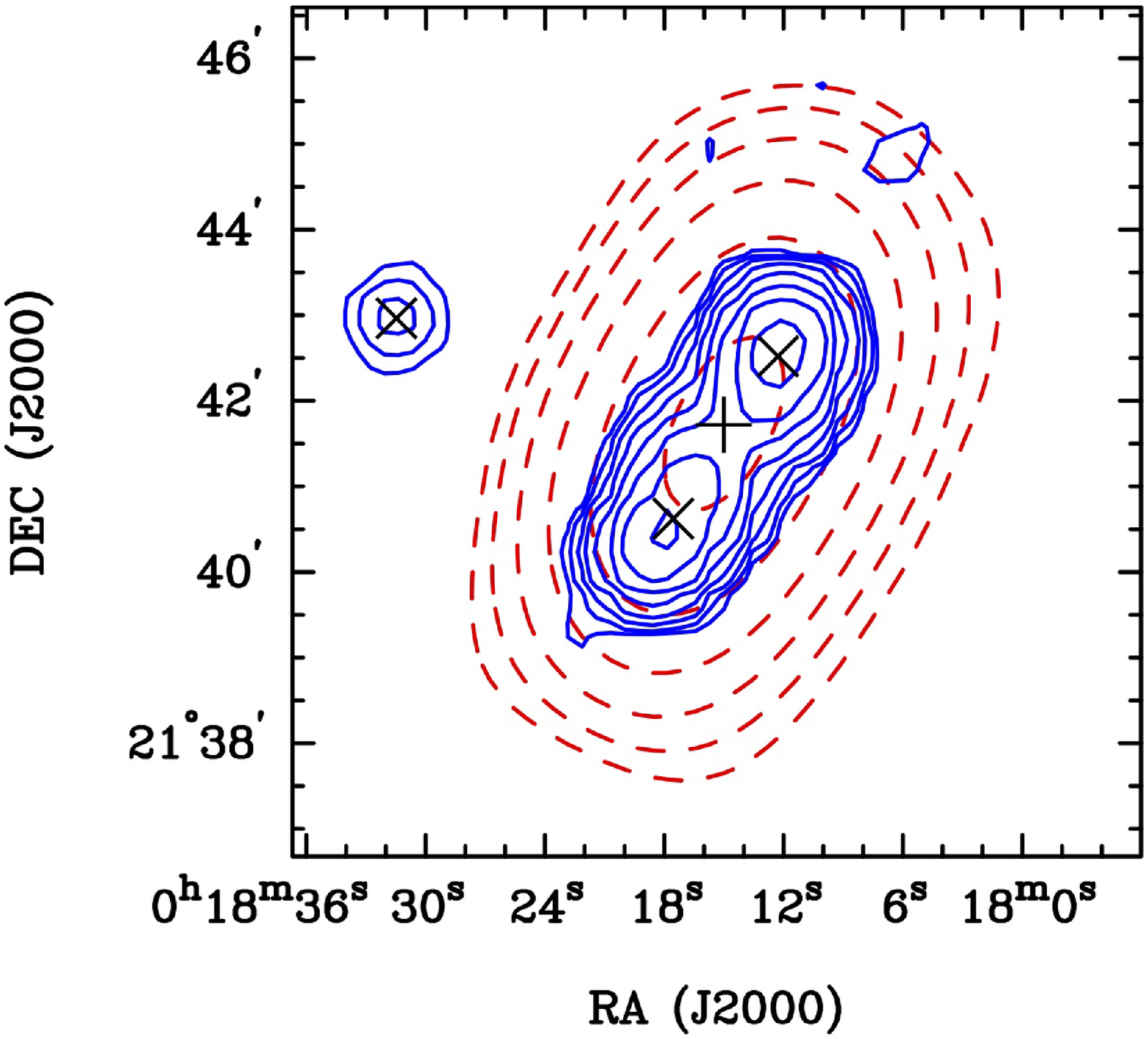}
\caption{Two GLEAM catalogued sources revealed as having double-lobed radio structure by the NVSS survey images: sources GLEAM J001636-382649 and GLEAM J001815+214143 respectively. Dashed contours (GLEAM) and solid contours (NVSS) are shown with the lowest contour level at 3$\sigma$ (rms, survey) and increasing in factors of 2. The catalogued radio positions are shown for both surveys, GLEAM $+$ and NVSS $\times$.}
\label{gdoubleb}
\end{figure}

Whilst the GLEAM survey has excellent surface brightness sensitivity, the image resolution varies as approximately 2.5 x 2.2/cos($\delta + 26.7^\circ$) arcmin at 151 MHz. Given these characteristics we determine that there are four instances to consider in translating the selected GLEAM sample to a physical source list:  (1) as will be described in Hurley-Walker et al. (in prep), GLEAM imaging has excised regions around the so-called `Class A' sources, e.g. Fornax A, etc, such that these sources are missing. These sources are well-determined and will be added into our 4 Jy sample at a later stage.  (2) Sources with two or more separately-catalogued GLEAM components with individual flux densities less than 4~Jy are not included in this sample: to quantify these we will use the 2-point angular correlation function $w(\theta)$ to estimate the instance of wide doubles. (3) Sources with very extended and separated radio lobes could be catalogued as individual components in GLEAM: where one or both lobes are brighter than 4~Jy the component(s) appear in our sample. These instances are resolved as a single radio source via visual inspection with other catalogue data as described later. (4) GLEAM sources with complex regions of low surface brightness: in this case often just one component appears in the selected sample. These instances are resolved as  described in the next section via analysis of GLEAM sources in close proximity coupled with visual inspection as also described below. Examples of each case are shown in Figures \ref{gdoublea} and \ref{gdoubleb}.

Having established how we interpret the GLEAM survey component catalogue to define a {\it complete} 4 Jy source sample, we refine the source positions using higher frequency, higher resolution, data. This step also provides a first insight to the nature of these sources.  We examine positional cross-matches of the GLEAM 4~Jy sources with existing higher frequency radio surveys which assists in resolving cases (3) and (4) noted above, as well as allowing us to refine the centroid position of the radio source emission of all GLEAM 4 Jy sources. 

We begin by analysing all 2130 selected GLEAM components. As noted in cases (3) and (4) above, whilst GLEAM has a large beam size it is still possible that the source finding process has fragmented sources; to find sources where this has occurred, we run a process to identify potentially related source components (`chains of friends') using a 4 arcmin step offset, i.e. approximately 2 beam widths.  We find that 1873/2130 sources are isolated GLEAM components having no near neighbours and the remaining 257 GLEAM components have one or more neighbours in close proximity.  On visual inspection it becomes clear that the vast majority of these close-proximity components are unrelated, such that (only) 27 of these are sources where a single 4~Jy radio source has been fragmented (i.e. fit with multiple Gaussian components) in GLEAM processing. We correct our sample by accumulating the flux densities of these multiple components and also calculate a new centroid-weighted GLEAM position for the complex source. 

The flux density distribution of the resultant 4 Jy sample reveals that it is dominated by sources at the faint end, as is expected given the steepness of the 151~MHz source counts. If we consider the equivalent 3CRR flux density limit at this frequency as 12.33 Jy (i.e. 10.9 Jy at 178 MHz transposed to 151~MHz assuming a spectral index of $-$0.75), then there are 1888 GLEAM 4~Jy sources with flux densities below this limit (89\% of the sample); thus $\sim$90\% of the 4Jy GLEAM sample will augment the information obtained from 3CRR. 


We cross-match the GLEAM 4~Jy source sample with the SUMSS (Bock, Large, Sadler 1999) and NVSS surveys (Condon et al. 1998) . Where there are multiple NVSS or SUMSS components detected within the GLEAM search radius we visually inspect the data and find 371 (17\%) are double, triple or higher order radio source structures, examples of which are shown in Figure 3.  

This step of cross-matching with the SUMSS and NVSS catalogues not only reveals information on the higher resolution, higher frequency, characteristics of the GLEAM 4 Jy sources but also helps us derive a more robust centroid position of the origin of the radio emission, via an adaption of the methodology developed by Magliocchetti et al. (1998). This work is ongoing and we will describe this process and the cross-match of this GLEAM sample to other waveband catalogues, including the ATCA 20 GHz survey (AT20G: Murphy et al. 2010), IR, optical, etc in a future publication (Jackson et al. in prep). 

\section{Discussion and future work}

As discussed at this conference, there are many opportunities for new insights into radio galaxy evolution and physical models from the upcoming generation of radio surveys. Our initial work described here outlines how we are using a first look at the brightest $\sim$2000 sources at 151~MHz to check the integrity of the GLEAM catalogue data, ahead of its public release, as well as our motivation in pursuing this sample to be completely-identified in the future. 


\section{Acknowledgements}

This work makes use of the Murchison Radio-astronomy Observatory, operated by CSIRO. We acknowledge the Wajarri Yamatji people as the traditional owners of the Observatory site. Support for the operation of the MWA is provided by the Australian Government Department of Industry and Science and Department of Education (National Collaborative Research Infrastructure Strategy: NCRIS), under a contract to Curtin University administered by Astronomy Australia Limited. We acknowledge the iVEC Petabyte Data Store and the Initiative in Innovative Computing and the CUDA Center for Excellence sponsored by NVIDIA at Harvard University. We acknowledge the International Centre for Radio Astronomy Research (ICRAR), a Joint Venture of Curtin University and the University of Western Australia, funded by the West Australian Government. CAJ thanks the Department of Science, Office of Premier \& Cabinet, WA 
for their support through the Western Australian Fellowship Program. The Centre for All-sky Astrophysics is an Australian Research Council Centre of Excellence, funded by grant CE110001020.

\end{document}